\documentclass[paper]{ieice}
\usepackage[dvipdfm]{graphicx}
\usepackage{graphicx}
\usepackage{latexsym}
\usepackage{amsmath}
\usepackage{amssymb}
\usepackage{bm}
\usepackage{amsmath,amssymb}
\usepackage{graphicx}

\newcommand{\pdif}[2]{\frac{\partial #1}{\partial #2}}
\newcommand{\coef}{\mathrm{coef}}
\newcommand{\zero}{\mathbf{0}}
\newcommand{\one}{\mathbf{1}}

\newcommand{\Rho}{\mathrm{P}}
\newenvironment{proof}{\noindent {\it Proof~:}\ }{\ \rule{1mm}{2mm}\medskip}

\newtheorem{theorem}{Theorem}
\newtheorem{lemma}{Lemma}

\vol{93}
\no{11}
\field{A}
\SpecialSection{Information Theory and Its Applications}
\title{Weight Distributions of Multi-Edge type LDPC Codes}
\authorlist{
 \authorentry[kenta@comm.ss.titech.ac.jp]{Kenta KASAI}{m}{TITECH}
 \authorentry[bee-t@comm.ss.titech.ac.jp]{Tomoharu AWANO}{n}{TITECH}
 \authorentry[declercq@ensea.fr]{David DECLERCQ}{n}{ENSEA}
 \authorentry[poulliat@ensea.fr]{Charly POULLIAT}{n}{ENSEA}
 \authorentry[sakaniwa@comm.ss.titech.ac.jp]{Kohichi SAKANIWA}{f}{TITECH}
}
\affiliate[TITECH]{Dept.\ of Communications and Integrated Systems, 
Tokyo Institute of Technology
2-12-1, O-Okayama, Meguro-ku, Tokyo, 152-8552, Japan}
\affiliate[ENSEA]{
ETIS ENSEA/University of Cergy-Pontoise/CNRS, F-95000, Cergy-Pontoise,
 Cergy, FRANCE
}

\received{2010}{2}{19}

\begin{document}
\maketitle
\begin{summary}
The multi-edge type LDPC codes, introduced by Richardson and Urbanke, present the  general class of structured LDPC codes.
In this paper, we derive the average weight distributions of the multi-edge type LDPC  code ensembles.
Furthermore, we investigate the asymptotic exponential growth rate of the average weight distributions and
investigate the connection to the stability condition of the density evolution.  
\end{summary}

\section{Introduction}
\label{223637_18Feb10}
In 1963, Gallager invented low-density parity-check (LDPC) codes \cite{gallager_LDPC}.
Due to the sparseness of the representation of the codes, 
LDPC codes are efficiently decoded by the sum-product (SP) decoders \cite{910572} or Log-SP decoders \cite{910577}. 
The Log-SP decoding is also known as the belief propagation.
By the powerful method {\it density evolution} \cite{910577}, invented by Richardson and Urbanke,
the messages of the Log-SP decoding are statistically evaluated. 
The optimized LDPC codes can realize the reliable transmissions at rate close to the Shannon limit \cite{richardson01design}.

Recently, many {\it structured} LDPC codes have been proposed:
accumulate repeat accumulate codes \cite{ara},
irregular repeat accumulate codes \cite{ira},
MacKay-Neal codes \cite{mn_code},
protograph codes \cite{protograph},
raptor codes \cite{raptor},
low-density generator-matrix codes \cite{ldgm} and so on.
These structured codes are usually designed for exploiting the structure to realize an 
excellent decoding performance, efficient encoding, a fast decoding algorithm, a parallel implementation and so on. 
Above all, the multi-edge type LDPC (MET-LDPC) codes \cite{met} give a general framework
that unifies all those structured LDPC codes.

The average weight distribution of codewords, which is simply referred to as {\itshape the average weight distribution},  
helps the analysis of the average performance of the maximum likelihood (ML) decoding \cite{bur_enum} and 
the typical minimum distance \cite{gallager_LDPC}. 
The decoding errors for the high SNR regions are mainly brought by codewords of small weight. 
Constructing LDPC codes without small weight codewords helps to lower the error floors. 
For the average weight distribution of the standard irregular LDPC codes are studied in \cite{di_wd,bur_enum,ipc,orlitsky_ss,mct} and 
those of structured codes are studied in \cite{coset,kasai_wd,ikegaya}.
Specifically for the standard irregular LDPC codes, the average weight distribution is derived in \cite{orlitsky_ss} as
the coefficients of some polynomial. And many useful properties \cite{di_wd,mct} are derived from the coefficient expression. 

In  \cite{kasai_wd}, we have already derived the average weight distributions for the MET-LDPC code ensembles. 
However, the derived equation  \cite{kasai_wd} is not as simple as that of the standard irregular LDPC codes \cite{orlitsky_ss}
and is hard to investigate further properties.
Indeed the derived equation in \cite{kasai_wd} is not written in a closed form 
but given as a recursive form which concatenates the weight distributions of 
constituent codes of the MET code ensembles.  
In this paper, we derive the average weight distributions of the MET-LDPC  code ensembles in a simple closed form.
Furthermore, we investigate the asymptotic exponential growth rate of the average weight distributions and
investigate the connection to the stability condition of the density evolution.  

The rest of this paper is organized as follows. 
Section \ref{223648_18Feb10} gives the definition of the MET-LDPC code ensemble. 
Section \ref{223728_18Feb10} gives the simple expression for the average weight distributions of the MET-LDPC code ensemble. 
In Section \ref{223808_18Feb10} we derive the asymptotic exponential growth rate of the average weight distributions 
and investigate it for the codeword of small weight in Section \ref{223908_18Feb10}. 
Furthermore, in Section \ref{223932_18Feb10},  
we show that the connection between the the asymptotic exponential growth rate of the codeword of small weight and 
the stability condition of the density evolution \cite{richardson01design}.  
\section{Multi-Edge type LDPC Codes}
    \label{223648_18Feb10}
\begin{figure}[t]
\begin{center}
  \includegraphics[scale=1.2]{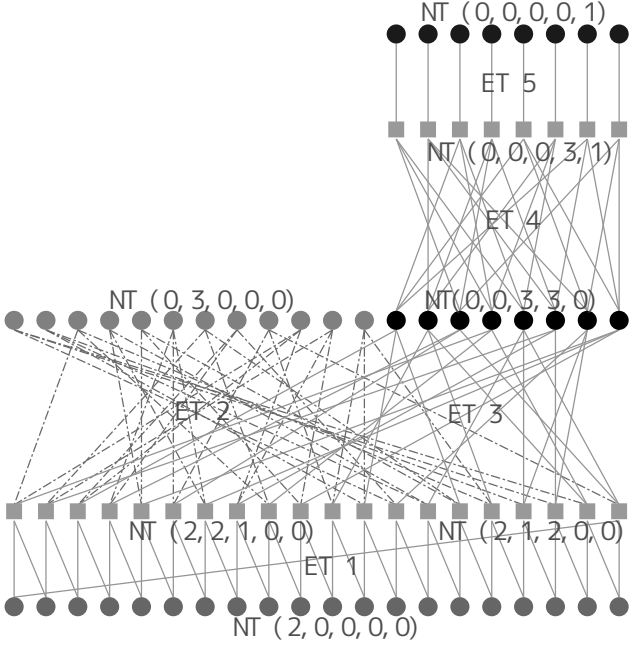}
\caption{The Tanner graph of a Multi-Edge type LDPC code. NT and ET stand for variable node-type, check node-type and edge-type, respectively. }
\label{fig:graph}
\end{center}
\end{figure}
Before we give the general definition of the ME-LDPC codes, we show a specific example of MET-LDPC code for a better understanding. 
The Tanner graph of an example of MET-LDPC code is shown  in Fig.~\ref{fig:graph}.
We use the terms, a ``code'' and its ``Tanner graph'' interchangeably. 
The edges in the graphs are divided into 5 types of edges labeled from ``ET 1'' to ``ET 5''. 
Note that
there are two types of edges in the third row of edges. 
With this classification, each edge is said to have {\itshape edge-type} $i$ for $i=1,\dotsc,5$.
Furthermore variable and check nodes are classified into types according
to the number of each edge-type they have. The types are called variable
and check {\itshape node-types}, respectively.
For example, the check nodes labeled ``NT(2,2,1,0,0)'' are said to have node-type (2,2,1,0,0) since 
these check nodes have two edges of edge-type 1, two edges of edge-type 2 and 1 edge of edge-type 3. 
And the variable nodes labeled ``NT(0,0,0,0,1)'' are said to have node-type (0,0,0,0,1) since 
these variable nodes have 1 edge of edge-type 5. 

The original definition of MET-LDPC \cite{met} codes involves the transmissions over
the $n_{\mathfrak r}$ types of parallel channels. Since our interest in this
paper is limited to the average weight distributions, we restrict ourselves to the
transmissions over a single channel, i.e. $n_{\mathfrak r}=1$.


For simplicity of notation, we define $\mathbf{1}=(1,\dotsc,1)$ and $\mathbf{0}=(0,\dotsc,0)$.
And define that $\mathbf{x}\ge\zero$ means that $x_i\ge 0$ for $i=1,\dotsc,n$. 
Moreover, we use the notation
\begin{align*}
 \mathbf{x}^{\mathbf{y}}:=\prod_{i=1}^{n}x_i^{y_i}
\end{align*}
for two vectors $\mathbf{x}=(x_1,\dotsc,x_n), \mathbf{y}=(y_1,\dotsc,y_n)$ of size $n$. 

Now, we give the definition of an MET-LDPC code ensemble. 
Analogously to the degree distribution pair ($\lambda(x), \rho(x)$) \cite{mct} for the standard irregular LDPC code ensemble, 
an MET-LDPC code ensemble is specified by a multivariate polynomial pair
$(\nu(\mathbf{r}, \mathbf{x}), \mu(\mathbf{x}))$ which is also referred to as {\itshape the degree distribution pair}.
\begin{align*}
& \nu(\mathbf{r}, \mathbf{x}):=\sum_{\mathbf{b}, \mathbf{d}\ge\zero}\nu_{\mathbf{b}, \mathbf{d}}\mathbf{r}^{\mathbf{b}}\mathbf{x}^{\mathbf{d}}, \\
&\mu(\mathbf{x}):=\sum_{\mathbf{d}\ge\zero}\mu_{\mathbf{d}}\mathbf{x}^{\mathbf{d}},\\
&\mathbf{b}:=(b_0, b_1, \ldots, b_{n_{\mathfrak r}}), \mathbf{d}:=(d_1,  \ldots, d_{n_{\mathfrak e}}),\\
&\mathbf{r}:=(r_0, r_1, \ldots, r_{n_{\mathfrak r}}), \mathbf{x}:=(x_1,  \ldots, x_{n_{\mathfrak e}})
\end{align*}
where $n_{\mathfrak r}$ is the number of channel-types and
$n_{\mathfrak e}$ is the number of edge-types.

For a given degree distribution pair $(\nu(\mathbf{r}, \mathbf{x}), \mu(\mathbf{x}))$ and code length $n$, 
we define an equi-probable ensemble of LDPC codes with graphs $G$ that satisfy the
followings. 
\begin{enumerate}
 \item
      $G$ has $n$ variable nodes of channel-type $\bm{b}=(0, 1)$, i.e. $G$ has
      $n$ un-punctured transmitted bits.
 \item
      $G$ has $n\nu_{\mathbf{b}, \mathbf{d}}$ variable nodes of channel-type
      $\mathbf{b}$ and node-type $\mathbf{d}$.
 \item
      $G$ has $n\mu_{\mathbf{d}}$ check nodes of node-type $\mathbf{d}$.
\end{enumerate}
We denote this code ensemble  by ${\mathcal C}(n, \nu(\mathbf{r}, \mathbf{x}), \mu(\mathbf{x}))$.

It is easy to see that the number of edges of edge-type $i$ incident to
variable and check nodes of node-type $\mathbf{d}$ are respectively given as
\begin{align*}
 &\sum_{\mathbf{b}\ge\zero}d_in\nu_{\mathbf{b}, \mathbf{d}}=d_in\nu_{(1,0), \mathbf{d}}+d_in\nu_{(0,1), \mathbf{d}},\\
 &\sum_{\mathbf{d}\ge\zero}d_i n\mu_{\mathbf{d}}.
\end{align*} 
It follows that the number of edges of edge-type $i$ incident to variable
nodes and check nodes are respectively given as
\begin{align*}
&n\nu_{i}({\mathbf 1}, {\mathbf 1}):=\left.n\frac{\partial}{\partial x_i}\nu({\mathbf r}, {\mathbf x})\right|_{\substack{\mathbf{r}=\mathbf{1}\\\mathbf{x}=\mathbf{1}}}
=n\sum_{\mathbf{b}, \mathbf{d}\ge\zero}d_i\nu_{\mathbf{b}, \mathbf{d}},\\
&n\mu_{i}({\mathbf 1}):=\left.n\frac{\partial}{\partial x_i}\mu({\mathbf x})\right|_{\mathbf{x}=\mathbf{1}}
=n\sum_{\mathbf{d}\ge\zero}d_i\mu_{\mathbf{d}},
\end{align*}
for $i=1,\dotsc,n_{\mathfrak e}$, where $\mathbf{1}:=(1,\dotsc,1)$.
They are constrained to be identical, and we denote this number by $E_i$, i.e. 
for $i=1,\dotsc, n_{\mathfrak{e}}$
\begin{align*}
 E_i:=n\nu_{i}({\mathbf 1}, {\mathbf 1})=n\mu_{i}({\mathbf 1}).
\end{align*}
Being permuted the connection among $E_i$ edges of edge-type $i$ in a graph, the resulting Tanner graph has the same degree distribution pair. 
The number of  graphs in the MET-LDPC ensemble is given as follows. 
\begin{align}
 \#{\mathcal C}(n, \nu(\mathbf{r}, \allowbreak\mathbf{x}),\allowbreak\mu(\mathbf{x})) =\prod_{i=1}^{n_\mathfrak{e}}E_i!.\label{120937_16Feb10}
\end{align}

In this setting, we can see that the graph shown in Fig.~\ref{fig:graph} is an MET-LDPC code in 
the MET-LDPC code ensemble $\mathcal{C}(n=40, \nu(\mathbf{r}, \mathbf{x}),
\mu(\mathbf{x}))$, where
 \begin{align*}
 \nu(\mathbf{r}, \mathbf{x})&=0.5r_1x_1^2+0.3r_1x_2^3+0.2r_0x_3^3x_4^3+0.2r_1x_5,\\
 \mu(\mathbf{x})&=0.4x_1^2x_2^2x_3+0.1x_1^2x_2x_3^2+0.2x_4^3x_5.
 \end{align*}
\section{Weight Distribution of Multi-Edge type Codes}
\label{223728_18Feb10}
In this section, we derive the average weight distribution
of the MET-LDPC code ensemble ${\mathcal C}(n, \nu(\mathbf{r}, \mathbf{x}), \mu(\mathbf{x}))$.
For readers who are unfamiliar with the enumeration technique of the weight distributions in LDPC code ensembles, we refer the readers to \cite{gallager_LDPC,bur_enum}.

We consider all the $2^N$ maps from each variable node to $\{0,1\}$, 
\begin{align*}
 x:v\mapsto x_v\in\{0,1\}. 
\end{align*}
We say a map $x$ is a codeword of a code $G$ if $\sum_{v\in V_c}x_v$ is an even number
for every check node $c$ in $G$, where $V_c$ is the set of variable nodes adjacent to the check node $c$ in $G$. 
The weight $w(x)$ of a map $x$ is defined as the number of un-punctured variable nodes $v$
such that $x_v=1$. 
Let $A_G(\ell)$ be the number of codewords of weight $\ell$ in a code $G$. 
Let $A(\ell)$ be the average number of codewords of weight $\ell$ for the MET-LDPC code
 ensemble ${\mathcal C}(n, \nu(\mathbf{r}, \allowbreak\mathbf{x}),\allowbreak\mu(\mathbf{x}))$ defined as follows. 
\begin{align*}
 A(\ell)=\sum_{G\in{\mathcal C}(n, \nu(\mathbf{r}, \allowbreak\mathbf{x}),\allowbreak\mu(\mathbf{x}))}A_G(\ell)\big/\#{\mathcal C}(n, \nu(\mathbf{r}, \allowbreak\mathbf{x}),\allowbreak\mu(\mathbf{x})). 
\end{align*}
\begin{theorem}
\label{theorem:A(l)}
For a given MET-LDPC code ensemble
${\mathcal C}(n, \nu(\mathbf{r}, \allowbreak\mathbf{x}),\allowbreak\mu(\mathbf{x}))$, 
the average number of codewords of weight $\ell$ 
is given as follows.
\begin{align}
A(\ell)=&\sum_{\mathbf{e}\ge\zero}
\frac{
\coef(
\left(
Q(t, \mathbf{s})
P(\mathbf{u})
\right)^n,
t^{\ell}\mathbf{s}^{\mathbf{e}}\mathbf{u}^{\mathbf{e}})}
{\prod_{i}\binom{E_i}{e_i}},\label{120109_16Feb10}\\
\nonumber Q(t, \mathbf{s})=&\prod_{\mathbf{b},
 \mathbf{d}\ge\zero}(1+t^{b_1}\mathbf{s}^{\mathbf{d}})^{\nu_{\mathbf{b}, \mathbf{d}}}, \\
\nonumber P(\mathbf{u})=&\prod_{\mathbf{d}\ge\zero}
\left(
\frac{(\mathbf{1}+\mathbf{u})^{\mathbf{d}}+(\mathbf{1}-\mathbf{u})^{\mathbf{d}}}{2}
\right)^{\mu_{\mathbf{d}}},
\end{align}
where
\begin{align*}
 & \mathbf{e}=(e_1,\dotsc,e_{n_{\mathfrak e}}),\\ 
 &\mathbf{u}=(u_1,\dotsc,u_{n_{\mathfrak e}}),\\
 &\mathbf{s}=(s_1,\dotsc,s_{n_{\mathfrak e}}).
\end{align*}
And $\coef(g(\mathbf{x}), \mathbf{x}^{\mathbf{d}})$ is the coefficient
of a term  $\mathbf{x}^{\mathbf{d}}$ in a multivariate
 polynomial $g(\mathbf{x})$.
\end{theorem}
\proof
An edge is said to be {\itshape active}, if the edge is incident to a
 variable node $v$ such that $x_v=1.$
We will count all the codewords of weight $\ell$ in all graphs in the ensemble 
${\mathcal C}(n, \nu(\mathbf{r}, \allowbreak\mathbf{x}),\allowbreak\mu(\mathbf{x}))$
with $e_i$
active edge of edge-type $i$ for $i=1,\dotsc, n_{\mathfrak{e}}$, and sum them up for all $\mathbf{e}=(e_1,\dotsc,e_{n_{\mathfrak e}})\ge \zero$. 
Counting all the codewords involves the following 3 parts: 
\begin{enumerate}
 \item Count the active edge constellations satisfying all the parity-check constraints.
 \item Count the active edge constellations which stem from maps of weight $\ell$.
 \item Count the edge permutations among active edges and non-active edges.
\end{enumerate}

Before we start counting the active edge constellations satisfying all the parity-check constraints, 
first, let us count the active edge constellations satisfying a single parity-check constraint. 
Consider a check node $c$ of node-type $\mathbf{d}$. 
In other words, the check node $c$ has $d_i$ edges of edge-type $i$ for $i=1,\dotsc,n_{\mathfrak{e}}$. 
The check node $c$ is satisfied if the total number  of active edges is even, i.e.~ $\sum_{i=1}^{n_{\mathfrak{e}}}e_i=\text{even}$. 
Let $a_\mathrm{c}(\mathbf{e})$  be the number of active edge constellations which satisfy the check node $c$ 
with given $e_i$ active incident edges of edge-type $i$ for $i=1, \ldots, n_{{\mathfrak e}}$.
It is easily checked that 
\begin{align*}
 a_\mathrm{c}(\mathbf{e})=
\left\{
\begin{array}{ll}
 \prod_{i=1}^{n_{\mathfrak{e}}}\binom{d_i}{e_i} & \sum_{i=1}^{n_{\mathfrak{e}}}e_i=\text{even}\\
 0 & \sum_{i=1}^{n_{\mathfrak{e}}}e_i=\text{odd}
 \end{array}
\right.
\end{align*}
Let $f_{\mathbf{d}}(\mathbf{u})$ the generating function of $a_\mathrm{c}(\mathbf{e})$ defined as
\begin{align*}
f_{\mathbf{d}}(\mathbf{u}):=\sum_{\mathbf{e}\ge \zero}a_\mathrm{c}(\mathbf{e})\mathbf{u^e}. 
\end{align*}
We can simply describe $f_{\mathbf{d}}(\mathbf{u})$ as
 \begin{align*}
f_{\mathbf{d}}(\mathbf{u})
&= \frac{\prod_{i=1}^{n_{\mathfrak{e}}}(1+u_i)^{d_i}+\prod_{i=1}^{n_{\mathfrak{e}}}(1-u_i)^{d_i}}{2}\\
&= \frac{(\one+\mathbf{u})^{\mathbf{d}}+(\one-\mathbf{u})^{\mathbf{d}}}{2}. 
 \end{align*}
Next, count the active edge constellations satisfying 
all the $n\mu(\mathbf{1})$ parity-check constraints with given $e_i$ active edges of edge-type $i$ for
$i=1, \ldots, n_{{\mathfrak e}}$. 
Since there are $n\mu_\mathbf{d}$ check nodes of node-type $\mathbf{d}$ for $\mathbf{d}\ge\zero$, 
the number of  active edge constellations to satisfy all the
parity-check constraints is given by
\begin{align}
\label{eq:coef_c}
  \coef(\prod_{\mathbf{d}\ge\zero}f_{\mathbf{d}}(\mathbf{u})^{n\mu_{\mathbf{d}}}, \mathbf{u}^{\mathbf{e}}).
\end{align}

Secondly, we will count the active edge constellations which stem from maps of weight $\ell$.
Count the active edge constellations which stem from a single variable node of weight $\ell=0,1$ at first. 
This may be somewhat confusing since it is too trivial. 
Consider a variable node $v$ of channel-type $\mathbf{b}$ and node-type $\mathbf{d}$. 
Assume $v$ is given $e_i$ active edges of edge-type $i$ for $i=1, \ldots, n_{{\mathfrak e}}$. 
Let $a_\mathrm{v}(\ell, \mathbf{b}, \mathbf{e})$  be the number of constellations which stem from 
the maps of weight $\ell\in\{0,1\}$.
From the definition of the active edges, it is easily checked that 
\begin{align*}
 a_\mathrm{v}(\ell, \mathbf{b}, \mathbf{e})=
\left\{
\begin{array}{ll}
 1 &(\ell=0, \mathbf{e}=\mathbf{0}),\\
 1 &(\ell=b_1, \mathbf{e}=\mathbf{d}),\\
 0 &\text{otherwise}.
 \end{array}
\right.
\end{align*}
Therefore, the generating function of $ a_\mathrm{v}(\ell, \mathbf{b}, \mathbf{e})$ is simply written as
\begin{align*}
 \sum_{\ell\in\{0,1\},\mathbf{b}\ge\zero,\mathbf{e}\ge\zero}a_{\mathrm{v}}(\ell,\mathbf{b},\mathbf{e})t^{\ell}\mathbf{s}^{\mathbf{e}}=1+t^{b_1}\mathbf{s^d}.
\end{align*}
Next, consider all $n$ variable nodes. 
There are $n\nu_\mathbf{b,d}$ variable nodes of channel-type $\mathbf{b}$ and of node-type $\mathbf{d}$ 
for $\mathbf{b}\ge\zero$ and $\mathbf{d}\ge\zero$. 
It is consequent  that for given $e_i$ active edges of edge-type $i$, the number of active edge constellations which stem from maps $x:v\mapsto x_v\in \{0,1\}$ of weight $\ell$
is given by
\begin{align}
\label{eq:coef_v}
 \coef(\prod_{\mathbf{b}, \mathbf{d}\ge\zero}(1+t^{b_1}\mathbf{s}^{\mathbf{d}})^{n\nu_{\mathbf{b}, \mathbf{d}}}, t^{\ell}\mathbf{s}^{\mathbf{e}}).
\end{align}

In the third place, 
consider that we are given $e_i$ active edges of edge-type $i$
and hence there are $E_i-e_i$ non-active edges of type $i$ 
for $i=1,\ldots, n_{{\mathfrak e}}$. 
For these active edge and non-active edges, 
the number of possible ways of permuting active and non-active edges is given as
\begin{align}
\label{eq:num_connecting_edge}
 \prod_{i=1}^{n_{\mathfrak{e}}}e_i!(E_i-e_i)!.
\end{align}

Let $A_{\mathbf{e}}(\ell)$ be 
the average number of graphs which have codewords of weight $\ell$ 
for given $e_i$ active edges of type $i$ for $i=1,\dotsc,n_{\mathfrak{e}}$. 
By multiplying Eq.~\eqref{eq:coef_c}, Eq.~\eqref{eq:coef_v} and
 Eq.~\eqref{eq:num_connecting_edge}, and
dividing by the number of codes in the ensemble given in Eq.~\eqref{120937_16Feb10},
we obtain 
 \begin{align*}
A_{\mathbf{e}}(\ell)=&\coef(\prod_{\mathbf{d}\ge\zero}f_{\mathbf{d}}(\mathbf{u})^{n\mu_{\mathbf{d}}}, \mathbf{u}^{\mathbf{e}})\\
&\cdot{\coef(\prod_{\mathbf{b}, \mathbf{d}\ge\zero}(1+t^{b_1}\mathbf{s}^{\mathbf{d}})^{n\nu_{b, \mathbf{d}}}, t^{\ell}\mathbf{s}^{\mathbf{e}})}
\Big/\prod_{i=1}^{n_\mathfrak{e}}\binom{E_i}{e_i}.
 \end{align*}
The average number of codewords of weight $\ell$ for the ensemble is obtained by 
summing up $A_{\mathbf{e}}(\ell)$ over the all possible active edge numbers.
 \begin{align}
 A(\ell)&=\sum_{\mathbf{e}\ge \zero} A_{\mathbf{e}}(\ell)\label{121218_16Feb10}
 \end{align}
 \QED
 \section{Asymptotic Analysis}
 \label{223808_18Feb10}
 LDPC codes are usually used with large code length. 
We are interested in the asymptotic average weight distributions in the limit of large code length.
The average number of codewords of weight $\omega n$  is usually increases or 
decays exponentially in $n$. 
We focus our interest in the asymptotic exponential growth rate of the $A(\ell)$ which is 
simply referred to as the {\itshape growth rate} $\gamma(\omega) $ defined as follows. 
\begin{align*}
  \displaystyle\gamma(\omega):=\lim_{n\to\infty}\frac{1}{n}\log A(\omega n), 
\end{align*}
where $\omega$ called  is the normalized weight of codewords.

In this section, we derive the growth rate for the MET-LDPC code ensemble. 
To this end, first introduce the following lemma.
 \begin{lemma}[\cite{bur_enum}, III.2]
\label{lemma:bur}
 For an $m$-variable polynomial $g(x_1, \dotsc, x_m)$ with non-negative coefficients,
  it holds that
  \begin{align*}
  \lim_{n\to\infty}\frac{1}{n}\log\coef(g(\mathbf{x})^n, \mathbf{x}^{\bm{\alpha}n})=\inf_{\mathbf{x}>\mathbf{0}}\log\frac{g(\mathbf{x})}{\mathbf{x}^{\bm{\alpha}}},
  \end{align*}
where $\mathbf{x}>\mathbf{0}$ means $x_i>0$ for all
  $i=1,\dotsc,m$. The point $\mathbf{x}$ that takes the minimum of $\frac{g(\mathbf{x})}{\mathbf{x}^{\bm{\alpha}}}$ is given
  by a solution of the following equations.
 \begin{align*}
 \frac{x_i}{g(\mathbf{x})}\frac{\partial g(\mathbf{x})}{\partial x_i}=\alpha_i \quad(i=1, 2, \dotsc, m)
 \end{align*}
 \end{lemma}

 The number of terms in  Eq.~\eqref{120109_16Feb10}
 is upper-bounded by $\prod_{i=1}^{n_{\mathfrak{e}}}E_i$. 
Therefore the largest term 
 alone contributes the growth rate of $A(\ell)$.
Therefore, from Eq.~\eqref{121218_16Feb10} we have
 \begin{align}
\max_{\mathbf{e}\ge \zero}A_{\mathbf{e}}(\ell) \le A(\ell)\le\big(\prod_{i=1}^{n_{\mathfrak{e}}}E_i\big) \max_{\mathbf{e}\ge \zero}A_{\mathbf{e}}(\ell)
 \end{align}
 \begin{align}
\frac{1}{n}\log A(\ell)= \frac{1}{n}\log\max_{\mathbf{e}\ge \zero}A_{\mathbf{e}}(\ell) + o(1). 
 \end{align}
Rewriting $A_{\mathbf{e}}(\ell)$ as 
\begin{align}
&A_{n\bm{\beta}}(\omega n)= 
\frac{
\coef(
\left(
Q(t, \mathbf{s})
P(\mathbf{u})
\right)^n,
(t^{\omega }\mathbf{s}^{\bm{\beta}\ge \zero}\mathbf{u}^{\bm{\beta}})^n)}
{\prod_{i=1}^{n_{\mathfrak e}}\binom{\mu_i(\mathbf{1})n}{\beta_in}},\nonumber\\
&\bm{\beta}=(\beta_1,\dotsc,\beta_{n_{\mathfrak{e}}}), \nonumber
\end{align}
where $\bm{e}=n\bm{\beta}$ and using Lemma \ref{lemma:bur}, we obtain that 
\begin{align}
\nonumber &\lim_{n\to\infty}\frac{1}{n}\log A(\ell)=
\sup_{\bm{\beta}\ge \zero}
\inf_{t>0, \mathbf{s}>\zero, \mathbf{u}>\zero}
\biggl[\\
\nonumber&\quad\log Q(\mathbf{s}, t)+\log P(\mathbf{u})-\sum_{i=1}^{n_{\mathfrak e}}\beta_i\log(u_i)\nonumber
\\
\nonumber &\quad-\sum_{i=1}^{n_{\mathfrak e}}\beta_i\log(s_i)-\omega\log(t)
 -\sum_{i=1}^{n_{\mathfrak e}}\mu_i(\mathbf{1})h\left(\frac{\beta_i}{\mu_i(\mathbf{1})}\right)
\biggr]\nonumber\\
 &=:
\sup_{\bm{\beta}\ge \zero}
\gamma(\bm{\beta})
\label{eq:sup_gamma}
\end{align}
A point $(\mathbf{u}, \mathbf{s}, t)$ that takes $\inf_{t, \mathbf{s},
\mathbf{u}}$ is given as a solution of the following equations.
\begin{align}
&\omega=\frac{t\pdif{Q}{t}}{Q}=\sum_{\mathbf{b}, \mathbf{d}\ge\zero}
\frac{\nu_{\mathbf{b},\mathbf{d}}b_1t\mathbf{s}^\mathbf{d}}
{1+t^{b_1}\mathbf{s}^\mathbf{d}},\label{eq:cnd_omega}\\
 &\beta_i=u_i\frac{\pdif{P}{u_i}}{P}=u_i\sum_{\mathbf{d}\ge\zero}\mu_\mathbf{d}d_i\frac
{\frac{(\mathbf{1}+\mathbf{u})^\mathbf{d}}{1+u_i}-\frac{(\mathbf{1}-\mathbf{u})^\mathbf{d}}{1-u_i}}
{\scriptstyle{(\mathbf{1}+\mathbf{u})^\mathbf{d}+(\mathbf{1}-\mathbf{u})^\mathbf{d}}},
\label{eq:cnd_u}\\
&\beta_i= s_i\frac{\pdif{Q}{s_i}}{Q}=
\sum_{\mathbf{b}, \mathbf{d}\ge\zero}\frac{\nu_{\mathbf{b}, \mathbf{d}} d_it^{b_1}\mathbf{s}^\mathbf{d}}{1+t^{b_1}\mathbf{s}^\mathbf{d}},
\text{ for } i=1,\dotsc,n_{\mathfrak{e}}.\label{eq:cnd_s}
\end{align}
A point $\bm{\beta}=(\beta_1, \dotsc, \beta_{n_{\mathfrak e}})$
which gives $\sup_{\bm{\beta}}$ needs to satisfy the stationary condition
\begin{align}
& \frac{\beta_i}{\mu_i(\mathbf{1})-\beta_i}={u_i}{s_i}
\label{eq:cnd_beta}.
\end{align}
Thus, we obtain the following theorem.
\begin{theorem}
\label{theorem:gamma}
For a given MET-LDPC code ensemble ${\mathcal C}(n, \nu(\mathbf{r}, \allowbreak\mathbf{x}),
 \allowbreak\mu(\mathbf{x}))$, the growth  rate 
 of the normalized weight $\omega$ is given by
\begin{align*}
 \displaystyle \gamma(\omega):=\lim_{n\to\infty}\frac{1}{n}A(\omega n)\nonumber
 =\max_{\bm{\beta}\in \mathfrak{B(\omega)}}\gamma(\bm{\beta}),
\end{align*}
where $\mathfrak{B}(\omega)$ is a set of $\bm{\beta}$ such that
 \eqref{eq:cnd_omega}, \eqref{eq:cnd_u}, \eqref{eq:cnd_s} and
 \eqref{eq:cnd_beta} hold.
\end{theorem}

The derivative of $\gamma(\bm{\beta})$ in terms of $\omega$ can be
 expressed in the following simple expression.
 \begin{lemma}
 \label{lemma:-log(t)}
 For $\bm{\beta}$ and $t$ such that  $t\neq 0$ and equations \eqref{eq:cnd_omega}, \eqref{eq:cnd_u} and
  \eqref{eq:cnd_s} hold, we have the following.
 \begin{align*}
  \frac{d}{d\omega}\gamma(\bm{\beta})=
 -\log(t(\omega))
 \end{align*}
 \end{lemma}
\proof
Let $x'$ denote the derivation of $x$ with respect to  $\omega$.
Differentiating $\gamma(\omega)$ defined in \eqref{eq:sup_gamma}, we have
\begin{align}
\label{eq:dgamma}
&\frac{d}{d\omega}\gamma(\bm{\beta})=
\frac{Q'}{Q}+\frac{P'}{P}
-w\frac{t'}{t}
-\sum_{i=1}^{n_{\mathfrak e}}
\log\frac{\mu_i(\mathbf{1})-\beta_i}{\beta_i}\beta_i'
\nonumber
\\
&
-\log t
-\sum_{i=1}^{n_{\mathfrak e}}(
\beta_i'\log u_i
+\beta_i\frac{u_i'}{u_i}
+\beta_i'\log s_i
+\beta_i\frac{s_i'}{s_i}),
\nonumber
\end{align}
where $\mathbf{s}$ is given by equations \eqref{eq:cnd_omega}, \eqref{eq:cnd_u} and
  \eqref{eq:cnd_s}. 
From \eqref{eq:cnd_beta}, we see
\begin{align*}
 -\beta_i'\log u_i-\beta_i'\log
 s_i-\beta_i'\log\frac{\mu_i(\mathbf{1})-\beta_i}{\beta_i}=0.
\end{align*}
Combining  \eqref{eq:cnd_u} and $ P' =\sum_{i=1}^{n_{\mathfrak{e}}}\pdif{P}{u_i}u_i'$, we have
\begin{align}
 \frac{P'}{P} -\sum_{i=1}^{n_{\mathfrak e}}\beta_i\frac{u_i'}{u_i}=0
\end{align}
From \eqref{eq:cnd_omega}, \eqref{eq:cnd_s} and $Q' =\pdif{Q}{t}t'+\sum_{i=1}^{n_{\mathfrak{e}}}\pdif{Q}{s_i}s_i'$, we have
\begin{align*}
\frac{Q'}{Q}-w\frac{t'}{t}+\sum_{i=1}^{n_{\mathfrak{e}}}\beta_i\frac{s_i'}{s_i}=0
\end{align*}
Thus, we can conclude the proof since the remaining term in
the right hand side of \eqref{eq:dgamma} is $-\log t$.
\QED
\section{Analysis of Small Weight Codeword}
\label{223908_18Feb10}
In this section we restrict ourselves to considering un-punctured MET-LDPC codes, i.e.
\begin{align}
\label{assumption:un-puncture}
 \mathbf{b}=(b_0,b_1)=(0, 1) \text{ for } \nu_{\mathbf{b}, \mathbf{d}}\neq 0.
\end{align}
Furthermore, we assume that for every edge-type $i$ there exists a check node
which has at least 2 edges of edge-type $i$.
In precise, for  $i=1,\dotsc,n_{\mathfrak{e}},$
\begin{align}
\label{assumption:check}
 \exists \mathbf{d}\text{ such that }d_i\ge 2   \text{ and } \mu_{\mathbf{d}}\neq 0.
\end{align}
For the standard irregular LDPC codes
\cite{luby01improved} with
a degree distribution pair $(\lambda(x), \rho(x))$, this assumption 
reduces to the condition of  the non-existence of 
check nodes of degree 1, i.e.~$\rho'(1)>0$.

We investigate how the growth rate 
behaves for codewords of small weight, i.e. for small normalized weight
$\omega$.
From the linearity of MET-LDPC codes, $A(0)=1$ and $\gamma(0)=0$,
then from \eqref{eq:sup_gamma} and  Lemma \ref{lemma:-log(t)}, it follows that
for $\omega\to 0$,
\begin{align}
\label{eq:gamma'(0)}
\gamma(\omega)&=\gamma'(0)\omega + o(\omega)\\
             & =\sup_{t\in \mathfrak{T}}-\log (t)\omega +o(\omega),
\end{align}
where $\mathfrak{T}$ is a set of $t$ such that 
\eqref{eq:cnd_omega}, \eqref{eq:cnd_u}, \eqref{eq:cnd_s} and
\eqref{eq:cnd_beta} hold
for $\omega\to 0$.
From the assumption of non-puncturing \eqref{assumption:un-puncture} and \eqref{eq:cnd_omega}, for $\omega\to 0$, it holds that
$t\mathbf{s}^\mathbf{d}\to 0$ for $\mathbf{d}$ with $\nu_{\mathbf{b}, \mathbf{d}}\neq 0$.
Using this, it follows that $\beta_i\to 0$ for $i=1,\dotsc,n_{\mathfrak{e}}$ from \eqref{eq:cnd_s}. 
Using  the assumption of check node-types Eq.~\eqref{assumption:check} and  Eq.~\eqref{eq:cnd_u}, 
it is consequent that $u_i\to 0$  for $i=1,\dotsc,n_{\mathfrak{e}}$.
Moreover, from \eqref{eq:cnd_u} it follows that as $\mathbf{u}\to\mathbf{0}$, 
 \begin{align*}
 \beta_i=\sum_{\mathbf{d}\ge \zero}\mu_{\mathbf{d}}u_id_i((d_i-1)u_i+\sum_{j\neq i}d_ju_j)+o(\textstyle{(\sum_{i=1}^{n_{\mathfrak{e}}}u_i)^2}).
 \end{align*}
Substituting this to \eqref{eq:cnd_beta}, we have
 \begin{align}
 \label{eq:s=Pu}
 & {s_i}=\frac{\mu_{i, i}(\mathbf{1})}{\mu_i(\mathbf{1})}u_i+\sum_{j\neq i}\frac{\mu_{i, j}(\mathbf{1})}{\mu_i(\mathbf{1})}u_j+o(\textstyle{\sum_{i=1}^{n_{\mathfrak{e}}}u_i}),\\
 &\mu_{i, j}(\mathbf{x})=\frac{\partial^2}{\partial x_i\partial x_j}\mu(\mathbf{x}).\nonumber
 \end{align}
As $\mathbf{s}\to \mathbf{0}$, from \eqref{eq:cnd_s} we have
 \begin{align*}
 \beta_i&=ts_i({\nu_{i, i}}(\mathbf{1}, \mathbf{0}) s_i+\sum_{j\neq
  i}\nu_{i, j}(\mathbf{1}, \mathbf{0})s_j) +o(\textstyle{(\sum_{i=1}^{n_{\mathfrak{e}}}s_i)^2})
 \end{align*}
Substituting this to \eqref{eq:cnd_beta}, we obtain the following.
\begin{align}
\label{eq:u=tLs}
{u_i} &=t\Bigl(\frac{{\nu_{i, i}}(\mathbf{1}, \mathbf{0})}{\nu_i(\mathbf{1}, \mathbf{1})}s_i
+\sum_{j\neq i}\frac{{\nu_{i, j}}(\mathbf{1},
 \mathbf{0})}{\nu_i(\mathbf{1}, \mathbf{1})}s_j\Bigr) +o(\textstyle{\sum_{i=1}^{n_{\mathfrak{e}}}s_i})\\
&\nu_{i, j}(\mathbf{r}, \mathbf{x})=\frac{\partial^2}{\partial x_i\partial x_j}\nu(\mathbf{r}, \mathbf{x})\nonumber
\end{align}
We can represent  \eqref{eq:s=Pu} and \eqref{eq:u=tLs} by matrices
 as  $\mathbf{s}=\Rho \mathbf{u}$ and $ \mathbf{u}=t\Lambda(\mathbf{1})
 \mathbf{s}$, respectively,
where
 \begin{align*}
& \Lambda_{i, j}(\mathbf{r}):=\frac{\left.\frac{\partial^2\nu(\mathbf{r}, \mathbf{x})}{\partial x_i\partial x_j}\right|_{\mathbf{x}=\mathbf{0}}}{\nu_i(\mathbf{1}, \mathbf{1})},\\
& \Rho_{i, j}:=\frac{\left.\frac{\partial^2\mu(\mathbf{x})}{\partial x_i\partial x_j}\right|_{\mathbf{x}=\mathbf{1}}}{\mu_i(\mathbf{1})}.
 \end{align*}
 In summary, for $t\neq 0$ we obtain
 \begin{align}
 \frac{1}{t}\mathbf{u}=\Lambda(\mathbf{1})
  \Rho\mathbf{u}+o(\textstyle{\sum_{i=1}^{n_{\mathfrak{e}}}u_i}).
 \end{align}
This implies that $\frac{1}{t}$ is an eigenvalue of $\Lambda(\mathbf{1})\Rho$.
Therefore $\sup_{t\in \mathfrak{T}}$ of \eqref{eq:gamma'(0)} is achieved by
the largest eigenvalue $\frac{1}{t}$ of $\Lambda(\mathbf{1}) \Rho$.
Then we have the following theorem.
 \begin{theorem}
 \label{theorem:met-gamma}
 For an MET-LDPC code ensemble ${\mathcal C}(n, \nu(\mathbf{r},
  \allowbreak\mathbf{x}), \allowbreak\mu(\mathbf{x}))$,
 assume  the largest eigenvalue $\frac{1}{t}$ of
  $\Lambda(\mathbf{1})\Rho$ is not zero. 
  The growth rate $\gamma(\omega):=\lim_{n\to\infty}\frac{1}{n}\log A(\omega n)$
  of the average number $A(\omega n)$ of codewords of weight $\omega n$,
  in the limit of code length, is given by
 \begin{align*}
 \gamma(\omega)=\log\left(\frac{1}{t}\right)\omega + O(\omega^2).
 \end{align*}
Furthermore, there exists $\delta > 0$ such that if $\frac{1}{t}<1$, 
there are exponentially few codewords of
weight $\omega n$ for $\omega < \delta$. 
 \end{theorem}

For a standard irregular LDPC code ensemble \cite{mct} with a given degree distribution pair $(\lambda(x), \rho(x))$,
can be viewed as an MET-LDPC code ensemble 
\begin{align*}
 {\mathcal C}\Bigl(n, \nu(r_1,x)= r_1\frac{\sum_i\lambda_i{x^i}/{i}}{\sum_i{\lambda_i}/{i}}, 
 \mu(x)=\frac{\sum_i\rho_i{x^i}/{i}}{\sum_i{\lambda_i}/{i}}\Bigr).
\end{align*}
The eigenvalue is given by $\lambda'(0)\rho'(1)$ which is zero if there
are no variable nodes of degree 2. The condition $\frac{1}{t}<1$ in Theorem
\ref{theorem:met-gamma} reduces
  to $\lambda'(0)\rho'(1)<1$, which coincides with the known result \cite{di_wd}. 
\section{Relation with Stability Condition}
\label{223932_18Feb10}
In this section, we investigate the connection between 
 the growth rate and  the stability condition \cite{richardson01design}.
For simplicity, we assume the transmission takes place over the binary erasure channels (BEC) with the erasure
probability $\epsilon$. 

For the standard irregular LDPC code ensemble \cite{mct} with degree distribution pair ($\lambda(x),\rho(x)$), 
in the limit of the code length, 
we denote the average decoding erasure probability of messages sent from variable nodes to check nodes at the $\ell$-th iteration round by $p^{(\ell)}$.
From density evolution \cite{richardson01design},  
$p^{(\ell)}$ is given by
 \begin{align*}
&p^{(0)}=\varepsilon,\\
&p^{(\ell)}=\varepsilon\lambda(1-\rho(1-p^{(\ell-1)})).
 \end{align*}
The following is shown in \cite{richardson01design},
if $\varepsilon\lambda'(0)\rho'(1)>1$, there exists $\gamma>0$ such that $\lim_{\ell\to\infty}p^{\ell}>\gamma$.
The inequality 
\begin{align}
 \varepsilon\lambda'(0)\rho'(1)<1\label{223544_19Feb10}
\end{align}
is called  {\itshape the stability condition of density evolution}.  
Furthermore, it is shown in \cite{1077796}, the capacity-achieving LDPC code ensemble have the degree distribution pair ($\lambda(x), \rho(x)$) with $\varepsilon\lambda'(0)\rho'(1)=1$. 
Meanwhile, it is known that the growth rate  of the average number
of codewords of small linear weight $\omega n$ is given by
\begin{align}
\lim_{n\to\infty}\frac{1}{n}\log A(\omega n)= \log(\lambda'(0)\rho'(1))\omega + o(\omega).\label{223551_19Feb10}
\end{align}
Interestingly,  the same parameter $\lambda'(0)\rho'(1)$ appears in
both the stability condition Eq.~\eqref{223544_19Feb10} and the growth rate Eq.~\eqref{223551_19Feb10}. 
Does this correspondence  also hold for the MET-LDPC code ensembles?

For the MET-LDPC code ensemble with the degree distribution pair $(\nu(\mathbf{r},\mathbf{x}), \allowbreak\mu(\mathbf{x}))$,
in the limit of large code length, 
let $p_i^{(\ell)}$ denote the erasure probability of the message sent along the edges of edge-type $i$ from variable nodes to check nodes at the $\ell$-th interation round.

From the density evolution developed for the MET-LDPC codes \cite[Eq.~(8)]{met}, 
$p_i^{(\ell)}$ is recursively given by
 \begin{align*}
\mathbf{p}^{(\ell)}&=\bm{\lambda}((1, \varepsilon),  \mathbf{1}-\bm{\rho}(\mathbf{1}-\mathbf{p}^{(\ell-1)})), \\
\bm{\lambda}(\mathbf{r}, \mathbf{x}):&=(\lambda_1(\mathbf{r}, \mathbf{x}),\dotsc,\lambda_{n_\mathfrak{e}}(\mathbf{r}, \mathbf{x})),\\
\bm{\rho}(\mathbf{x}):&=(\rho_1(\mathbf{x}),\dotsc,\rho_{n_\mathfrak{e}}(\mathbf{x})),\\
 \lambda_i(\mathbf{r}, \mathbf{x})&=\frac{\nu_i(\mathbf{r}, \mathbf{x})}{\nu_i(\mathbf{1}, \mathbf{1})},\\
 \rho_i(\mathbf{x})&=\frac{\mu_i(\mathbf{x})}{\mu_i(\mathbf{1})},
  \end{align*}
where
 $p_i^{(0)}=\varepsilon$ for $i=1,\dotsc, n_{\mathfrak e}$.
And it follows that \cite[Theorem 7]{met} is given as follows.
If the spectral radius of $\Lambda(1, \varepsilon) \Rho$ is less than
 1, there exists $\gamma>0$ such that 
\begin{align*}
 \lim_{\ell\to\infty}\sum_ip_i^{(\ell)}>\gamma.
\end{align*}
In short, the stability condition for the MET-LDPC code is   given as follows. 
\begin{align*}
 1 >\text{the spectral radius of } \Lambda(1, \varepsilon) \Rho.
\end{align*}
Since  $\Lambda(1, \varepsilon) \Rho$ is a non-negative matrix,
the spectral radius of $\Lambda(1, \varepsilon) \Rho$ is an eigenvalue
 of $\Lambda(1, \varepsilon) \Rho$, it follows $\Lambda(1, 1) \Rho$
 coincides with the parameter which appears in Theorem \ref{theorem:met-gamma}.
 \section{Conclusion}
We present a simple expression of the average weight distributions of MET-LDPC
code ensembles which gives us a general framework of LDPC
codes. We showed that the correspondence between the growth rate of the weight
distributions and the stability condition is also the case with the MET-LDPC
codes.
\bibliographystyle{ieicetr}
\bibliography{IEEEabrv,kenta_bib,ryutaroh_bib}

\begin{thebibliography}{10}

\bibitem{gallager_LDPC}
R.G. Gallager, Low {D}ensity {P}arity {C}heck {C}odes, in Research Monograph
  series, MIT Press, Cambridge, 1963.

\bibitem{910572}
F.~Kschischang, B.~Frey, and H.A. Loeliger, ``Factor graphs and the sum-product
  algorithm,'' {IEEE} Trans. Inf. Theory, vol.47, no.2, pp.498--519, Feb.\
  2001.

\bibitem{910577}
T.~Richardson and R.~Urbanke, ``The capacity of low-density parity-check codes
  under message-passing decoding,'' {IEEE} Trans. Inf. Theory, vol.47, no.2,
  pp.599--618, Feb.\ 2001.

\bibitem{richardson01design}
T.J. Richardson, M.A. Shokrollahi, and R.L. Urbanke, ``Design of
  capacity-approaching irregular low-density parity-check codes,'' {IEEE}
  Trans. Inf. Theory, vol.47, pp.619--637, Feb.\ 2001.

\bibitem{ara}
A.~Abbasfar, D.~Divsalar, and K.~Yao, ``Accumulate-repeat-accumulate codes,''
  {IEEE} Trans. Commun., vol.55, no.4, pp.692--702, April\ 2007.

\bibitem{ira}
A.K. H.~Jin and R.~Mc{E}liece, ``Irregular repeat-accumulate codes,'' Proc. 2nd
  Int. Symp. on Turbo Codes and Related Topics, pp.1--8, Sept.\ 2000.

\bibitem{mn_code}
D.J. MacKay and R.M. Neal, ``Good error-correcting codes based on very sparse
  matrices,'' 1999.

\bibitem{protograph}
J.~Thorpe, ``Low-density parity-check ({LDPC}) codes constructed from
  protographs,'' IPN Progress Report, pp.42--154, Aug.\ 2003.

\bibitem{raptor}
A.~Shokrollahi, ``Raptor codes,'' {IEEE} Trans. Inf. Theory, vol.52, no.6,
  pp.2551--2567, June\ 2006.

\bibitem{ldgm}
J.~Cheng and R.~Mc{E}liece, ``Some high-rate near capacity codecs for the
  gaussian channel,'' Proc. 34th Annual Allerton Conf. on Commun., Control and
  Computing, Oct.\ 1996.

\bibitem{met}
T.~Richardson and R.~Urbanke, ``Multi-edge type {LDPC} codes,'' 2003.

\bibitem{bur_enum}
D.~Burshtein and G.~Miller, ``Asymptotic enumeration methods for analyzing
  {LDPC} codes,'' {IEEE} Trans. Inf. Theory, vol.50, no.6, pp.1115--1131, June\
  2004.

\bibitem{di_wd}
C.~Di, T.~Richardson, and R.~Urbanke, ``Weight distribution of low-density
  parity-check codes,'' {IEEE} Trans. Inf. Theory, vol.52, no.11,
  pp.4839--4855, Nov.\ 2006.

\bibitem{ipc}
M.~Mezard and A.~Montanari, Information, Physics, and Computation, Oxford
  Graduate Texts, Oxford University Press, 2009.

\bibitem{orlitsky_ss}
A.~Orlitsky, K.~Viswanathan, and J.~Zhang, ``Stopping set distribution of
  {LDPC} code ensembles,'' {IEEE} Trans. Inf. Theory, vol.51, no.3,
  pp.929--953, March\ 2005.

\bibitem{mct}
T.~Richardson and R.~Urbanke, Modern Coding Theory, Cambridge University Press,
  March\ 2008.

\bibitem{coset}
T.~Wadayama, ``Average coset weight distribution of combined {LDPC} matrix
  ensembles,'' {IEEE} Trans. Inf. Theory, vol.52, no.11, pp.4856--4866, Nov.\
  2006.

\bibitem{kasai_wd}
K.~Kasai, Y.~Shimoyama, T.~Shibuya, and K.~Sakaniwa, ``Average coset weight
  distribution of multi-edge type {LDPC} code ensembles,'' IEICE Trans. Fundam.
  Electron. Commun. Comput. Sci., vol.E89-A, no.10, pp.2519--2525, Oct.\ 2006.

\bibitem{ikegaya}
R.~Ikegaya, K.~Kasai, Y.~Shimoyama, T.~Shibuya, and K.~Sakaniwa, ``Weight and
  stopping set distributions of two-edge type {LDPC} code ensembles,'' IEICE
  Trans. Fundam. Electron. Commun. Comput. Sci., vol.E88-A, no.10,
  pp.2745--2761, Oct.\ 2005.

\bibitem{luby01improved}
M.~Luby, M.~Mitzenmacher, A.~Shokrollahi, and D.~Spielman, ``Improved
  low-density parity-check codes using irregular graphs,'' {IEEE} Trans. Inf.
  Theory, vol.47, no.2, pp.585--598, Feb.\ 2001.

\bibitem{1077796}
P.~Oswald and A.~Shokrollahi, ``Capacity-achieving sequences for the erasure
  channel,'' {IEEE} Trans. Inf. Theory, vol.48, no.12, pp.3017 -- 3028, Dec.\
  2002.

\end{thebibliography}


\end{document}